\documentclass[aps,prd,onecolumn,groupedaddress,showpacs,nofootinbib,amssymb
]{revtex4}
\usepackage[dvips]{graphicx}
\usepackage{amssymb}
\usepackage{amsmath}
\usepackage{graphicx,color}
\usepackage{amsfonts}
\usepackage{bm}

\begin{document}

\tolerance=5000

\title{Effects of a Geometrically Realized Early Dark Energy Era on the Spectrum of Primordial Gravitational Waves}
\author{V.K. Oikonomou,$^{1}$}
\email{v.k.oikonomou1979@gmail.com,voikonomou@auth.gr,voiko@physics.auth.gr}
\author{Eirini C. Lymperiadou$^{1}$}
\email{eilymper@auth.gr}
\affiliation{$^{1)}$ Department of
Physics, Aristotle University of Thessaloniki, Thessaloniki 54124,
Greece}

\begin{abstract}
In this work we investigate the effects of a geometrically
generated early dark energy era on the energy spectrum of the
primordial gravitational waves. The early dark energy era, which
we choose it to have a constant equation of state parameter $w$,
is synergistically generated by an appropriate $f(R)$ gravity in
the presence of matter and radiation perfect fluids. As we
demonstrate, the predicted signal for the energy spectrum of the
$f(R)$ primordial gravitational waves is amplified and can be
detectable, for various reheating temperatures, especially for
large reheating temperatures. The signal amplitude depends on the
duration of the early dark energy era and on the value of the dark
energy equation of state parameter, with the most latter affecting
more crucially the amplification. Specifically the amplification
occurs when the equation of state parameter approaches the de
Sitter value $w=-1$. Regarding the duration of the early dark
energy era, we find that the largest amplification occurs when the
early dark energy era commences at a temperature $T=0.85\,$eV
until $T=7.8\,$eV. Moreover we study a similar scenario in which
amplification occurs, where the early dark energy era commences at
$T=0.29\,$eV and lasts until the temperature is increased by
$\Delta T\sim 1.7\,$eV.
\end{abstract}

\maketitle

\section*{Introduction}

One of the solutions proposed in the literature for solving the
$H_0$-tension problem is the existence of an early dark energy era
\cite{Niedermann:2020dwg,Poulin:2018cxd,Karwal:2016vyq,Oikonomou:2020qah,Nojiri:2019fft}.
If such an era is observed, then this could eliminate the
$H_0$-tension problem. In this work the focus is on the effects of
an early dark energy era on the energy spectrum of the primordial
gravitational waves generated by an $f(R)$ gravity. Specifically
we shall assume that an $f(R)$ gravity is the underlying geometric
source that drives inflation, the early dark energy era and the
late-time acceleration era, the commonly known dark energy era. In
our approach, inflation will be controlled by a vacuum $R^2$
gravity, while the early and ordinary dark energy era will
synergistically be controlled by appropriate $f(R)$ gravity terms
and radiation and cold dark matter fluids. Our aim is to see in a
quantitative way the direct effect of a geometrically generated
early dark energy era on the primordial gravitational waves energy
spectrum. For our analysis, we shall use the theoretically
predicted energy spectrum, which appears in many studies, see for
example Refs.
\cite{Kamionkowski:2015yta,Denissenya:2018mqs,Turner:1993vb,Boyle:2005se,Zhang:2005nw,Schutz:2010xm,Sathyaprakash:2009xs,Caprini:2018mtu,
Arutyunov:2016kve,Kuroyanagi:2008ye,Clarke:2020bil,Kuroyanagi:2014nba,Nakayama:2009ce,Smith:2005mm,Giovannini:2008tm,
Liu:2015psa,Zhao:2013bba,Vagnozzi:2020gtf,Watanabe:2006qe,Kamionkowski:1993fg,Giare:2020vss,Kuroyanagi:2020sfw,Zhao:2006mm,
Nishizawa:2017nef,Arai:2017hxj,Bellini:2014fua,Nunes:2018zot,DAgostino:2019hvh,Mitra:2020vzq,Kuroyanagi:2011fy,Campeti:2020xwn,
Nishizawa:2014zra,Zhao:2006eb,Cheng:2021nyo,Nishizawa:2011eq,Chongchitnan:2006pe,Lasky:2015lej,Guzzetti:2016mkm,Ben-Dayan:2019gll,
Nakayama:2008wy,Capozziello:2017vdi,Capozziello:2008fn,Capozziello:2008rq,Cai:2021uup,Cai:2018dig,Odintsov:2021kup,Benetti:2021uea,Lin:2021vwc,Zhang:2021vak,Odintsov:2021urx,Pritchard:2004qp,Zhang:2005nv,Baskaran:2006qs,Oikonomou:2022xoq,Odintsov:2022cbm,Odintsov:2022sdk}
and references therein, and we shall evaluate numerically the
effects of the $f(R)$ gravity on the general relativistic
waveform. We shall use a WKB approach developed in
\cite{Nishizawa:2017nef} and we shall investigate the amount of
amplification caused by an early dark energy era, which is
generated by an $f(R)$ gravity in the presence of perfect matter
and radiation fluids. The early dark energy era is described by a
constant equation of state (EoS) parameter $w$, and we shall find
which $f(R)$ gravity can generate such an evolution in the
presence of matter and radiation perfect fluids. Then we shall
calculate the overall amplification factor of the general
relativistic energy spectrum of the primordial gravitational
waves. Our results will be confronted with the sensitivity curves
of future interferometric experiments, like the LISA laser
interferometer space antenna \cite{Baker:2019nia,Smith:2019wny},
the DECIGO \cite{Seto:2001qf,Kawamura:2020pcg}, the Einstein
Telescope \cite{Hild:2010id}, the future BBO (Big Bang
Observatory) \cite{Crowder:2005nr,Smith:2016jqs}, and also the
non-interferometric experiments Square Kilometer Array (SKA)
\cite{Bull:2018lat} and the NANOGrav collaboration
\cite{Arzoumanian:2020vkk,Pol:2020igl}. As we shall demonstrate,
the general relativistic energy spectrum is amplified by the
presence of an $f(R)$ gravity generated early dark energy era,
only if several conditions hold true. Specifically, the duration
of the early dark energy era plays an important role and also the
value of the EoS parameter $w$ also crucially affects the results.
As we show, in most of the cases we studied, the energy spectrum
signal is amplified due to this $f(R)$ gravity generated early
dark energy era. As a conclusion, we point out that a detection of
a stochastic gravitational wave signal in future interferometric
experiments may have many possible explanations, and the early
dark energy realized by an f(R) gravity is one of these
explanations.

\section{$f(R)$ Gravity Realization of Inflation and Subsequent Eras}

The standard approach for realizing various cosmological eras in
Einstein-Hilbert cosmology is usually done by using perfect matter
fluids. The latter dominate the evolution at certain point, and
the corresponding era is controlled by those fluids, for example
the radiation domination era is controlled by the radiation fluid,
the energy density of which redshifts as $\rho_r\sim a^{-4}$,
where $a$ is the scale factor of a flat Friedmann-Robertson-Walker
(FRW) metric. Also for the matter domination era the matter
perfect fluid dominates the evolution, which describes
non-baryonic non-relativistic matter and its energy density
redshifts as $\rho_m\sim a^{-3}$. Apart from the three standard
evolution eras which we usually assume that the Universe
underwent, that is, the inflationary, radiation and matter
domination and dark energy era, we basically do not know the
behavior of our Universe post-inflationary. We have hints for the
post-inflationary era, but no proofs. This era is a mysterious
era, and it commences with the reheating era, which is the
beginning of radiation era, and it is believed that the reheating
era smoothly deforms to the radiation era. Remarkably, we also
lack of knowledge of what happened from the reheating era up to
the matter domination era, and before the recombination era. So
the question is what do we know? We know very well the physics
beyond the recombination era, up to the present day. The
recombination era is basically where the last scattering surface
of the CMB photons was formed, and from this era until present day
we understand the physics relatively well.

Thus the epoch before the recombination era we have no measured
data from that era, only at last scattering and beyond. So let us
assume that an early dark energy era is realized before the
recombination and specifically from the matter-radiation equality
redshift $z\sim 3400$ until some final redshift in the past,
deeply in the matter domination era $z_f$. In this work we shall
assume that this final redshift is a free variable and we shall
investigate what would be the effect of an early dark energy era
on the spectrum of the primordial gravitational waves, focusing on
modes which were subhorizon modes immediately after the
inflationary era and during the first stages of the reheating era.
Also, regarding the early dark energy era itself, we shall assume
that it is described by a constant EoS parameter, and more
importantly, it is not realized by some perfect matter fluid, but
it is realized geometrically, by some dominant form of $f(R)$
gravity for the whole early dark energy era. Also it is natural to
assume that $f(R)$ controls the inflationary and the late dark
energy era, as follows,
\[
f(R)=\left\{
\begin{array}{ccc}
  R+\frac{R^2}{6M^2}& R\sim R_I\, ,  \\
  F_w(R) &  R_{f}\leq R\ll R_{eq}\, ,   \\
  R+F_{DE}(R) & R\sim R_0\ll R_{rec}\, , \\
\end{array}\right.
\]
with $R_I$ being the curvature scale of inflation, which is
calculated primordially when the modes exit the horizon at the
first time, so at the beginning of inflation. Also the curvature
scale $R_{rec}$ is the curvature scale at the recombination,
$R_{f}$ is the curvature scale when the early dark energy era
commences, $R_{eq}$ the curvature at matter radiation equality and
$R_0$ is the curvature scale at present day, so it is basically
identical with the cosmological constant. Note that according to
our scenario, the era between the curvature scales $R_I<R<R_f$ is
not described by modified gravity. So modified gravity affects
post-inflationary  the evolution after $R>R_f$.

The exact forms of $F_w(R)$ and $F_{DE}(R)$ will be specified
shortly on the basis of phenomenological viability. Regarding the
function $F_w(R)$, this in conjunction with the matter and
radiation perfect fluids will synergistically generate the early
dark energy era with a constant EoS parameter $w$, and we shall
find that shortly. The function $F_{DE}(R)$ will realize the dark
energy era, at late-times, so some appropriate form of this will
be used in order to provide a viable dark energy era
phenomenology. In order to find the exact forms of the $f(R)$
gravity, we consider the gravitational action with perfect matter
fluids present,
\begin{equation}\label{action1dse}
\mathcal{S}=\frac{1}{2\kappa^2}\int
\mathrm{d}^4x\sqrt{-g}\,f(R)+\mathcal{S}_m\, ,
\end{equation}
with $\kappa^2=8\pi G=\frac{1}{M_p^2}$, with being $G$ Newton's
constant, and $M_p$ stands for the reduced Planck mass. In the
metric formalism, the field equations are,
\begin{equation}\label{eqnmotion}
f_R(R)R_{\mu \nu}(g)-\frac{1}{2}f(R)g_{\mu
\nu}-\nabla_{\mu}\nabla_{\nu}f_R(R)+g_{\mu \nu}\square
f_R(R)=\kappa^2T_{\mu \nu}^{m}\, ,
\end{equation}
with $T_{\mu \nu}^m$ being the energy momentum tensor of the
matter and radiation perfect fluids, and furthermore
$f_R=\frac{\mathrm{d}f}{\mathrm{d}R}$. For a flat FRW spacetime,
in which case the line element reads,
\begin{equation}
\label{metric} \centering {\rm d}s^2=-{\rm
d}t^2+a(t)^2\sum_{i=1}^{3}{({\rm d} x^{i})^2}\, ,
\end{equation}
the field equations become,
\begin{equation}\label{frwf1}
-18\left (4H(t)^2\dot{H}(t)+H(t)\ddot{H}(t)\right)f_{RR}(R)+3
\left(H^2(t)+\dot{H}(t)
\right)f_R(R)-\frac{f(R)}{2}+\kappa^2\left(\rho_m+\rho_r
\right)=0\, ,
\end{equation}
with $\rho_m$, $\rho_r$ standing for the cold dark matter energy
density and radiation energy density respectively.

Now let us proceed to the core of our analysis and we assume that
the Universe goes through an intermediate early dark energy era,
with a constant EoS parameter $w$, thus $p_{eff}=w\rho_{eff}$,
where $p_{eff}$ and $\rho_{eff}$ are the Universe's total pressure
and energy density. The early dark energy era will be assumed to
last from $z=3400$, so from the matter radiation equality, until a
redshift $z_{f}$ which will be a free parameter in our analysis.
During the early dark energy era, lasting from $z=3400$ up to
$z=z_f$, the scale factor of the Universe is,
\begin{equation}\label{scalefactorquasidesitter}
a(t)=a_{end}\,t^{\frac{2}{3(1+w)}}\, ,
\end{equation}
with $a_{end}$ being the scale factor at the redshift $z=z_f$. It
is easy to find which geometrical $f(R)$ theory can realize the
cosmology (\ref{scalefactorquasidesitter}) synergistically in the
presence of matter and radiation perfect fluids. In order to do so
we shall apply the formalism of Ref. \cite{Nojiri:2009kx}, so we
use the $e$-foldings number as a dynamical variable, so we have,
\begin{equation}\label{efoldpoar}
e^{-N}=\frac{a_i}{a}\, ,
\end{equation}
where $a_i$ is some initial value of the scale factor. Using the
$e$-foldings number $N$, the Friedmann equation takes the form,
\begin{equation}
\label{newfrw1} -18\left [ 4H^3(N)H'(N)+H^2(N)(H')^2+H^3(N)H''(N)
\right ]f_{RR}(R)+3\left [H^2(N)+H(N)H'(N)
\right]f_R(R)-\frac{f(R)}{2}+\kappa^2\rho=0\, ,
\end{equation}
where $\rho=\rho_m+\rho_r$. Introducing the auxiliary function,
$G(N)=H^2(N)$, the Ricci scalar is written as follows,
\begin{equation}\label{riccinrelat}
R=3G'(N)+12G(N)\, .
\end{equation}
Eventually the Friedman equation becomes,
 \begin{equation}
\label{newfrw1modfrom} -9G(N(R))\left[ 4G'(N(R))+G''(N(R))
\right]f_{RR}(R) +\left[3G(N)+\frac{3}{2}G'(N(R))
\right]f_R(R)-\frac{f(R)}{2}+\kappa^2\left(\rho_m+\rho_r
\right)=0\, ,
\end{equation}
where $G'(N)=\mathrm{d}G(N)/\mathrm{d}N$ and
$G''(N)=\mathrm{d}^2G(N)/\mathrm{d}N^2$. By solving the above
equation, one obtains the $f(R)$ gravity which realizes the given
scale factor of interest, which in our case is that of Eq.
(\ref{scalefactorquasidesitter}). Let us now proceed to find the
explicit form of the $f(R)$ gravity which realizes the scale
factor (\ref{scalefactorquasidesitter}). For the scale factor
(\ref{scalefactorquasidesitter}), $G(N)$ becomes,
\begin{equation}\label{gnfunction}
G(N)=\frac{4 e^{-3 N (w+1)}}{9 (w+1)^2}\, ,
\end{equation}
and we took $a_r=1$ for simplicity. Upon combining Eqs.
(\ref{riccinrelat}) and (\ref{gnfunction}), we get,
\begin{equation}\label{efoldr}
N=\frac{\log \left(\frac{4 (1-3 w)}{3 R (w+1)^2}\right)}{3 (w+1)}.
\end{equation}
The above in conjunction with the following,
\begin{equation}\label{mattenrgydens}
\rho_{tot} =\sum_i\rho_{i0}a_0^{-3(1+w_i)}e^{-3N(R)(1+w_i)}\, ,
\end{equation}
affect the Friedmann equation (\ref{newfrw1modfrom}), which
becomes,
\begin{align}
\label{bigdiffgeneral1} &a_1
R^2\frac{\mathrm{d}^2f(R)}{\mathrm{d}R^2}
+a_2R\frac{\mathrm{d}f(R)}{\mathrm{d}R}-\frac{f(R)}{2}+\sum_iS_{i}R^{
\frac{3(1+w_i) }{3(1+w)}}=0\, ,
\end{align}
where the index ``i'' takes values $i=(r,m)$ with $i=r$ indicating
radiation and $i=m$ indicating cold dark matter perfect fluids.
Also the parameters $a_1$ and $a_2$, $S_i$ and $A$ are defined in
the following way,
\begin{equation}
\label{apara1a2} a_1=\frac{3(1+w)}{4-3(1+w)},\,\,\,
a_2=\frac{2-3(1+w)}{2(4-3(1+w))},\,\,\,S_i=\frac{\kappa^2\rho_{i0}a_0^{-3(1+w_i)}}{[3A(4-3(1+w))]^{\frac{3(1+w_i)}{3(1+w)}}},\,\,\,A=\frac{4}{3(w+1)}
\end{equation}
By solving (\ref{bigdiffgeneral1}) we will get the $f(R)$ gravity
which generates the early dark energy era, which was denoted
$F_w(R)$ previously, so the solution is,
\begin{equation}
\label{newsolutionsnoneulerssss} F_{w}(R)=\left
[\frac{c_2\rho_1}{\rho_2}-\frac{c_1\rho_1}{\rho_2(\rho_2-\rho_1+1)}\right]R^{\rho_2+1}
+\sum_i
\left[\frac{c_1S_i}{\rho_2(\delta_i+2+\rho_2-\rho_1)}\right]
R^{\delta_i+2+\rho_2}-\sum_iB_ic_2R^{\delta_i+\rho_2}+c_1R^{\rho_1}+c_2R^{\rho_2}\,
,
\end{equation}
where $c_1,c_2$ are simple integration constants, and also
$\delta_i$ and $B_i$ are,
\begin{equation}
\label{paramefgdd}
\delta_i=\frac{3(1+w_i)-23(1+w)}{3(1+w)}-\rho_2+2,\,\,\,B_i=\frac{S_i}{\rho_2\delta_i}
\, ,
\end{equation}
with $i=(r,m)$. Let us now specify the late time era, and we shall
choose a convenient $f(R)$ gravity which is known to provide a
viable late-time phenomenology. We shall choose the one used in
Ref. \cite{Odintsov:2021kup} which is known to be a viable dark
energy $f(R)$ gravity model, which is,
\begin{equation}\label{starobinsky}
F_{DE}(R)=-\gamma \Lambda
\Big{(}\frac{R}{3m_s^2}\Big{)}^{\delta}\, ,
\end{equation}
where $m_s$ is $m_s^2=\frac{\kappa^2\rho_m^{(0)}}{3}$, and also
$\rho_m^{(0)}$ denotes the cold dark matter energy density at
present day. Furthermore, the parameter $\delta $ is chosen
$0<\delta <1$, $\gamma$ is an arbitrary dimensionless parameter,
and $\Lambda$ is the cosmological constant. Hence, the Universe's
evolution is controlled during inflation by an $R^2$ model, while
from $z_f$ and up to $z=3400$ by $F_w(R)$ given in Eq.
(\ref{newsolutionsnoneulerssss}), and at late times by $F_{DE}(R)$
given in Eq. (\ref{starobinsky}). In principle it is not hard to
find such a phenomenological model, for example such a
phenomenological model would look like,
\begin{align}\label{mainfreffective}
& f(R)=e^{-\frac{\Lambda }{R}}\left(R+\frac{R^2}{6 M^2}
\right)+e^{-\frac{R}{\Lambda }}\tanh
\left(\frac{R-R_{f}}{\Lambda}\right)\left(\frac{R-R_{eq}}{\Lambda}
\right) F_{DE}(R) +e^{-\frac{R}{\Lambda }}\tanh
\left(\frac{R-R_0}{\Lambda} \right)\left(F_w(R)-R-\frac{R^2}{6
M^2} \right)\, .
\end{align}
It should be noted that the above model is not the only one that
can reproduce the phenomenology we want to describe, it is one
example, but certainly not the only one. We just quote one example
for completeness. With regard to the early dark energy era we
shall consider three cases for values of the EoS parameter,
described below,
\begin{align}\label{scenarios}
&\mathrm{Scenario}\,\,I\,:\,\,\mathrm{EoS}\,\,w=-0.99\,\,\, ,\\
\notag &
\mathrm{Scenario}\,\,II\,:\,\,\mathrm{EoS}\,\,w=-\frac{1}{3}-0.001\,
.,
\\
\notag & \mathrm{Scenario}\,\,III\,:\,\,\mathrm{EoS}\,\,w=-0.7\, .
\end{align}
So Scenarios I and II basically describe the limiting cases of
accelerating expansion nearly a de Sitter one (Scenario I) and
nearly accelerating (Scenario II slightly smaller EoS parameter
compared to the value of the EoS parameter for which
non-accelerating nor decelerating cosmology occurs, namely
$w=-\frac{1}{3}$). Finally, for Scenario III the EoS parameter
takes an intermediate value for the sake of completeness.

Let us now proceed to the evaluation of the observational indices
relevant to the calculation of the energy spectrum of the
primordial gravitational waves. Specifically, we shall calculate
the tensor-to-scalar ratio and the tensor spectral index. We shall
be interested in modes with , which is the pivot scale used in
Planck. For $f(R)$ gravity, the tensor spectral index is
\cite{reviews1,Odintsov:2020thl,Odintsov:2021kup},
\begin{equation}\label{tensorspectralindexr2ini}
n_T\simeq -2\epsilon_1^2\, ,
\end{equation}
and the tensor-to-scalar ratio is
\cite{reviews1,Odintsov:2020thl,Odintsov:2021kup},
\begin{equation}\label{tensotoscalarratio}
r=48\epsilon_1^2\, ,
\end{equation}
with $\epsilon_1$ being the first slow-roll index
$\epsilon_1=-\dot{H}/H^2$. For the $R^2$ gravity we have
$\epsilon_1\simeq \frac{1}{2N}$, hence,
\begin{equation}\label{r2modeltensorspectralindexfinal}
n_T\simeq -\frac{1}{2N^2}\, ,
\end{equation}
and the corresponding tensor-to-scalar ratio is,
\begin{equation}\label{tensortoscalarfinal}
r=\frac{12}{N^2}\, .
\end{equation}
In the next section we shall evaluate numerically the energy
spectrum of the primordial gravitational waves at present day, for
all the modes that became subhorizon during the early stages of
the reheating era. We shall be interested in short wavelength
modes with $\lambda \gg 10\,$Mpc, or equivalently with wavenumbers
$k>10^{7}\,$Mpc$^{-1}$ up to $k>10^{18}\,$Mpc$^{-1}$, which
corresponds to the frequency range $10^{-7}<f<10^3\,$Hz.


\section{Primordial Gravitational Wave Energy Spectrum: The effects of an Early Dark Energy Phase}

In the next ten years, several space interferometers will provide
observational data on whether the theoretically predicted
stochastic gravitational wave background exists or not. Already in
the literature, the theoretical predictions for the stochastic
background of primordial gravitational waves are intensively
studied, see Refs.
\cite{Kamionkowski:2015yta,Denissenya:2018mqs,Turner:1993vb,Boyle:2005se,Zhang:2005nw,Schutz:2010xm,Sathyaprakash:2009xs,Caprini:2018mtu,
Arutyunov:2016kve,Kuroyanagi:2008ye,Clarke:2020bil,Kuroyanagi:2014nba,Nakayama:2009ce,Smith:2005mm,Giovannini:2008tm,
Liu:2015psa,Zhao:2013bba,Vagnozzi:2020gtf,Watanabe:2006qe,Kamionkowski:1993fg,Giare:2020vss,Kuroyanagi:2020sfw,Zhao:2006mm,
Nishizawa:2017nef,Arai:2017hxj,Bellini:2014fua,Nunes:2018zot,DAgostino:2019hvh,Mitra:2020vzq,Kuroyanagi:2011fy,Campeti:2020xwn,
Nishizawa:2014zra,Zhao:2006eb,Cheng:2021nyo,Nishizawa:2011eq,Chongchitnan:2006pe,Lasky:2015lej,Guzzetti:2016mkm,Ben-Dayan:2019gll,
Nakayama:2008wy,Capozziello:2017vdi,Capozziello:2008fn,Capozziello:2008rq,Cai:2021uup,Cai:2018dig,Odintsov:2021kup,Benetti:2021uea,Lin:2021vwc,Zhang:2021vak,Odintsov:2021urx,Pritchard:2004qp,Zhang:2005nv,Baskaran:2006qs,Oikonomou:2022xoq,Odintsov:2022cbm,Odintsov:2022sdk}
and references therein.
\begin{figure}
\centering
\includegraphics[width=22pc]{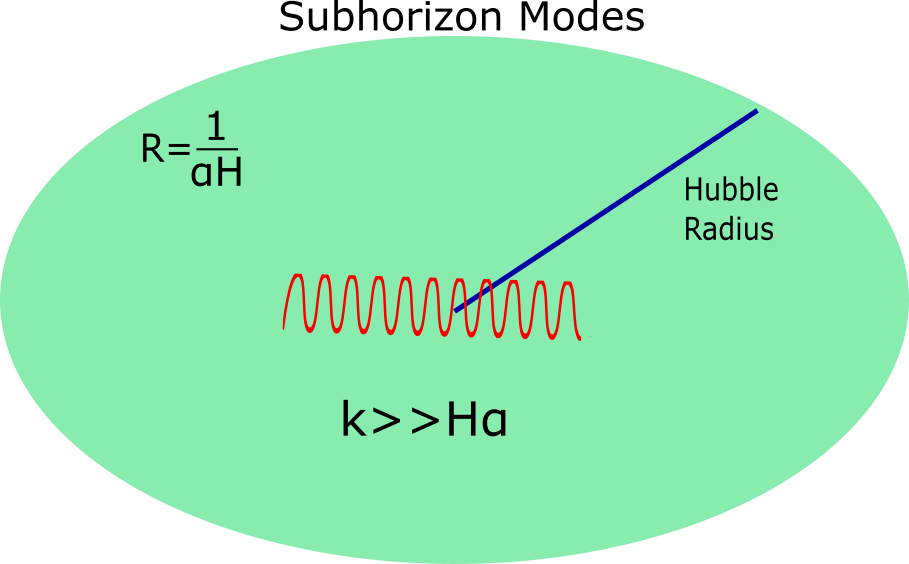}
\caption{Subhorizon inflationary modes which during the
inflationary era, these were at subhorizon scales, and have the
smallest wavelength. These are the first that reenter the horizon
after the end of the inflationary era and during the early stages
of the reheating era. The subhorizon modes will be probed by
future space interferometers.}\label{plot0}
\end{figure}
In this paper, the focus is to investigate the effect of an $f(R)$
gravity realized early dark energy era, with constant EoS
parameter $w$, which commences at the matter-radiation equality
and stretches up to a final redshift $z_f$, which shall be a free
variable for the moment. Our main assumption is that
$z_f\sim\mathcal{O}(10^4)$, but we shall allow for other higher
values for completeness, in order to see the effect on the energy
spectrum of the primordial gravitational waves. Let us note that
the choice of $z_f$ is arbitrary, it just indicates the end of the
early dark energy era. This is the maximum redshift which we will
allow the upper redshift limit of the early dark energy era to be.

\begin{figure}[h!]
\centering
\includegraphics[width=40pc]{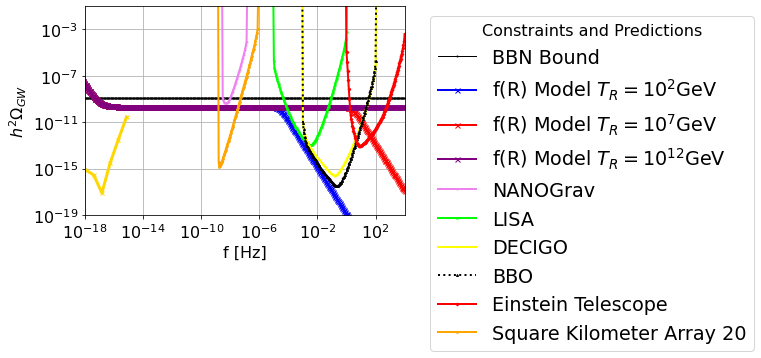}
\caption{The $h^2$-scaled gravitational wave energy spectrum for
the early dark energy $w$-era generated by $f(R)$ gravity, for the
Scenario I for $z_f=3\times 10^4$, with reheating temperatures
$T_R=10^{12}$GeV (purple), to $T_R=10^{7}$GeV (red) and to (blue)
$T_R=10^{2}$GeV.} \label{plot2}
\end{figure}

\begin{figure}[h!]
\centering
\includegraphics[width=40pc]{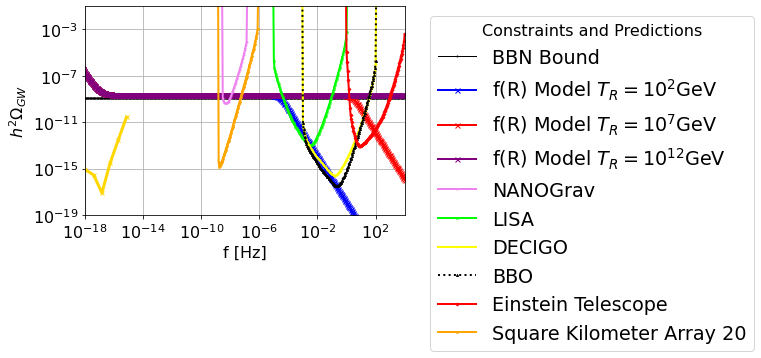}
\caption{The $h^2$-scaled gravitational wave energy spectrum for
the early dark energy $w$-era generated by $f(R)$ gravity, for the
Scenario I for $z_f=\times 10^4$ and for starting redshift of the
early dark energy era $z=1100$, with reheating temperatures
$T_R=10^{12}$GeV (purple), to $T_R=10^{7}$GeV (red) and to (blue)
$T_R=10^{2}$GeV.} \label{plot3}
\end{figure}

Let us discuss more the duration of the early dark energy era, and
as we stated we shall assume that it starts around the
matter-radiation equality at redshift $z=3400$ so at a temperature
$T=0.85\,$eV \cite{Garcia-Bellido:1999qrp} and it ends at $z_f$
which will be assumed in the range $z_f=[10^4,3\times 10^4]$ which
corresponds to a temperature range $T_f=[0.89,7.8]\,$eV
\cite{Garcia-Bellido:1999qrp}. In terms of the temperature, the
early dark energy era does not go deeply in the radiation
domination era, and specifically we shall assume that it lasts
from $T=0.89\,$eV up to $T_f=2.6\,$eV, however we shall extend the
duration up to $T_f=7.8\,$eV.
\begin{figure}[h!]
\centering
\includegraphics[width=40pc]{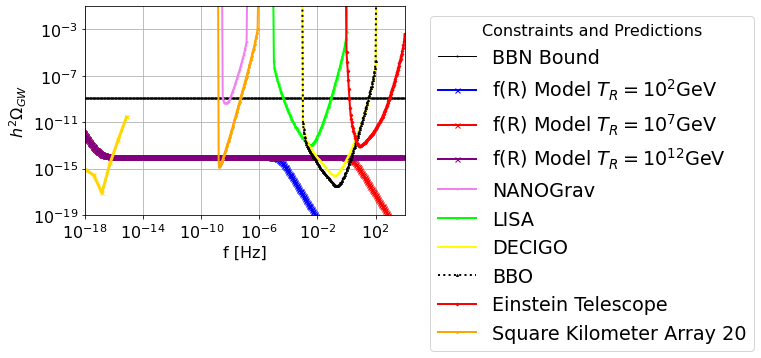}
\caption{The $h^2$-scaled gravitational wave energy spectrum for
the early dark energy $w$-era generated by $f(R)$ gravity, for the
Scenario I  for $z_f=10^5$, with reheating temperatures
$T_R=10^{12}$GeV (purple), to $T_R=10^{7}$GeV (red) and to (blue)
$T_R=10^{2}$GeV.} \label{plot1}
\end{figure}
The procedure to extract the overall effect of the $f(R)$ gravity
on the primordial gravitational waves is based on a WKB method
which applies on the modes that became subhorizon just after the
inflationary era ended, so during the reheating era. For a
pictorial representation of the subhorizon modes see Fig.
\ref{plot0}. The WKB method was developed in Refs.
\cite{Nishizawa:2017nef,Arai:2017hxj} and relies on the
calculation of the parameter $\alpha_M$, defined as,
\begin{equation}\label{amfrgravity}
a_M=\frac{f_{RR}\dot{R}}{f_RH}\, ,
\end{equation}
and the modified gravity effect on the waveform is,
\begin{equation}\label{mainsolutionwkb}
h=e^{-\mathcal{D}}h_{GR}\, ,
\end{equation}
where $h_{GR}$ is the waveform with $a_M=0$ which corresponds to
the general relativity case, and also the quantity $\mathcal{D}$
is defined as,
\begin{equation}\label{dform}
\mathcal{D}=\frac{1}{2}\int^{\tau}a_M\mathcal{H}{\rm
d}\tau_1=\frac{1}{2}\int_0^z\frac{a_M}{1+z'}{\rm d z'}\, .
\end{equation}
The primordial gravity waves energy density including the $f(R)$
gravity effects is
\cite{Boyle:2005se,Nishizawa:2017nef,Arai:2017hxj,Nunes:2018zot,Liu:2015psa,Zhao:2013bba,Odintsov:2021kup},
\begin{align}
\label{GWspecfR}
    &\Omega_{\rm gw}(f)=e^{-2\mathcal{D}}\times \frac{k^2}{12H_0^2}r\mathcal{P}_{\zeta}(k_{ref})\left(\frac{k}{k_{ref}}
\right)^{n_T} \left ( \frac{\Omega_m}{\Omega_\Lambda} \right )^2
    \left ( \frac{g_*(T_{\rm in})}{g_{*0}} \right )
    \left ( \frac{g_{*s0}}{g_{*s}(T_{\rm in})} \right )^{4/3}  \left (\overline{ \frac{3j_1(k\tau_0)}{k\tau_0} } \right )^2
    T_1^2\left ( x_{\rm eq} \right )
    T_2^2\left ( x_R \right )\, ,
\end{align}
with $k_{ref}=0.002$$\,$Mpc$^{-1}$ being CMB pivot scale. Also
$n_T$ denotes the tensor spectral index and $r$ denotes the
tensor-to-scalar ratio. The calculation of the quantity
$\mathcal{D}$ is the main aim hereafter, for the redshift ranges
$z=[0,3400]$, and $z=[3400,3\times 10^4]$ and to see the overall
amplification or damping on the energy spectrum. Thus the quantity
that needs to be calculated is,
\begin{equation}\label{dformexplicitcalculation}
\mathcal{D}=\frac{1}{2}\left(\int_0^{3400}\frac{a_{M_1}}{1+z'}{\rm
d z'}+\int_{3400}^{z_f}\frac{a_{M_2}}{1+z'}{\rm d z'}\right)\, ,
\end{equation}
with $a_{M_1}$ and $a_{M_2}$ being calculated for $f(R)\sim
R+f_{DE}(R)$ and $f(R)\sim F_w(R)$ respectively and with $z_f$
varying in the range $z_f=[10^4,3\times 10^4]$. Let us now present
our analysis for the three different scenarios we defined in the
previous section. For the scenarios II and III, the overall
amplification factors $\mathcal{D}$ are nearly zero, hence the
predicted energy spectrum of the primordial gravitational waves is
undetectable. Just for the sake of being precise, for scenario II,
the parameter $\mathcal{D}$ is $\mathcal{D}\sim 10^{-68}$ and
these results apply for both the redshift ranges $z=[3400, 10^4]$
and $z=[3400,3\times 10^4]$. Also the first integral
$\frac{1}{2}\left(\int_0^{3400}\frac{a_{M_1}}{1+z'}{\rm d
z'}\right)$ for all scenarios is of the order $\sim -0.05$ for all
the scenarios. The only non-trivial result occurs only for the
Scenario I, in which case we get
$\int_{3400}^{10^4}\frac{a_{M_2}}{1+z'}{\rm d z'}=-4.6353$ and
$\int_{3400}^{3\times 10^4}\frac{a_{M_2}}{1+z'}{\rm d
z'}=-9.35798$, so in both cases amplification occurs for the
energy spectrum. In Fig. \ref{plot1} and \ref{plot2} we plot the
$f(R)$ gravity $h^2$-scaled energy spectrum as a function of the
frequency, and specifically Fig. \ref{plot1} corresponds to the
case for which the early dark energy era lasts up to $z_f=10^4$
and Fig. \ref{plot2} for $z_f=3\times 10^4$. In both the plots we
also included the sensitivity curves of most of the future
interferometer experiments, and the predicted energy spectrum is
presented for three distinct reheating temperatures, for
$T_R=10^{12}$GeV (purple curve), for $T_R=10^{7}$GeV (red curve)
and $T_R=10^{2}$GeV (blue curve).
As it is obvious from both Figs. \ref{plot1} and \ref{plot2}, the
signals corresponding to the $f(R)$ gravity models are detectable
from most future experiments. Specifically, for the case with
$z_f=10^4$, the $f(R)$ gravity signal will be detected by the SKA
experiment for all reheating temperatures, and by the DECIGO and
BBO experiments for reheating temperatures $T_R=10^7$GeV and
$T_R=10^{12}$GeV. As for the case with $z_f=3\times 10^4$, in
which case the early dark energy era lasts slightly longer
compared to the standard scenarios in the literature, the signal
will be detected by all the experiments, except for the case with
reheating temperature $T_R=10^2$GeV, which will be detected only
by the SKA and NANOGrav experiments. Also let us note that if the
early dark energy era commences earlier, for example at the
recombination with $z\sim 1100$ and lasts until $z=10^4$, similar
results are obtained. Indeed in this case we get
$\int_{1100}^{10^4}\frac{a_{M_2}}{1+z'}{\rm d z'}=-10$, and in
Fig. \ref{plot3} we plot the $h^2$-scaled energy spectrum for this
case. As it can be seen the signal will be detected by all the
future experiments, except for the low reheating temperature
scenario. As an overall comment for all the cases we studied in
this section, it seems that the signal of $f(R)$ gravity
gravitational waves depends on the duration of the early dark
energy era, and for a good possibility of detection, the eras
between which the early dark energy era occurs must have a
temperature difference of approximately of the order $\Delta T\sim
1.7\,$eV. Also we need to comment on an important issue: since our
WKB approach affects the subhorizon modes during reheating, the
figures we presented must be looked with caution for small
frequencies, because these modes were superhorizon during
reheating. Hence our results are valid for frequencies starting
from the NANOGrav until the Einstein telescope.
Before closing this section, an important comment is in order. In
standard contexts of the Starobinsky inflationary scenario, the
signal of the primordial gravitational waves is undetectable. This
has to do with the fact that the tensor spectral index for the
standard Starobinsky inflation is negative, and thus, the energy
spectrum of the primordial gravitational waves generated is
significantly lower than the sensitivity curves of most future
gravitational waves experiments. This result however is mainly
based on a standard post-inflationary evolution, which includes
the reheating, radiation domination and matter domination eras.
However, with this work we aimed to show in a quantitative way
that a geometrically generated early dark energy era which occurs
just before recombination can amplify the energy spectrum
significantly. By geometrically generated, we mean that the origin
of this era is an $f(R)$ gravity in the presence of matter and
radiation perfect fluids. The reason for this amplification is an
overall amplifying factor $\sim e^{-2\mathcal{D}}$ appearing in
front of Eq. (\ref{GWspecfR}). This factor takes into account the
WKB effects of the modified gravity beyond the inflationary era.
The gravitational waves still redshift as radiation but their
energy spectrum is amplified, and this amplification is a direct
modified gravity effect, due to the fact that the evolution
equation contains a WKB overall factor. In most studies known for
the Starobinsky inflation predictions for primordial gravitational
waves, this post-inflationary amplification is absent, due to the
fact that the post-inflationary evolution is comprised by the
reheating, radiation domination and matter domination eras. In our
case, post-inflationary there are non-trivial effects generated by
the $f(R)$ gravity geometrically generated early dark energy era.
To show in a simple and brief way this issue, in the presence of a
non-trivial modified gravity, the evolution equation for tensor
perturbations is described by the following equation, of the
tensor perturbation $h_{i j}$,
\begin{equation}\label{mainevolutiondiffeqnfrgravity}
\ddot{h}(k)+\left(3+a_M \right)H\dot{h}(k)+\frac{k^2}{a^2}h(k)=0\,
,
\end{equation}
where $\alpha_M$ is in our case given in Eq. (\ref{amfrgravity}).
If post-inflationary the evolution is the standard one, then the
solution to the above equation is described by the one appearing
in Eq. (\ref{mainsolutionwkb}) without the factor $\sim
e^{-\mathcal{D}}$, thus $h(k)\sim h_{GR}$. In our case though, the
amplifying factor is non-trivial and as we showed numerically by
evaluating $\mathcal{D}$ in Eq. (\ref{dform}) for a constant EoS
parameter early dark energy era generated by $f(R)$ gravity in the
presence of matter and radiation perfect fluids, the amplification
is significant.

Furthermore, we need to clarify that the epochs before the early
dark energy era are not affected at all in our framework. The only
effect of the early dark energy era is contained in the amplifying
factor $\mathcal{D}$ in Eq. (\ref{GWspecfR}). The
post-inflationary epochs in between inflation and the early dark
energy era are contained in the Eq. (\ref{GWspecfR}), and even the
era of BBN is contained there. The BBN bounds are also taken into
consideration in Figs. \ref{plot1}-\ref{plot3}, corresponding to a
straight solid black line.

\section{Conclusions}

In this work we investigated quantitatively the effect of a
geometrically generated early dark energy era on the energy
spectrum of the primordial gravitations waves. Specifically the
early dark energy era is generated by an appropriate $f(R)$
gravity in the presence of matter and radiation perfect fluids.
The main assumption we made is that $f(R)$ gravity generates the
inflationary era, and specifically an $R^2$ gravity, the late-time
acceleration era, and also the early dark energy era. As we
showed, the energy spectrum of the primordial gravitational waves
is significantly amplified in a detectable way, and the signal can
be detected by most of the future gravitational waves experiments
that will seek for stochastic gravitational waves. Our analysis
indicated that the amplification of the signal depends on two
crucial parameters, the value of the dark energy EoS parameter $w$
and the duration of the early dark energy era. As we showed, the
amplification occurs only when the EoS parameter of the early dark
energy era is close to the de Sitter value and specifically we
studied the case $w=-0.9$. In addition the duration of the early
dark energy era affects the amplification. We assumed that the
early dark energy era started at $z=3400$ up to $z=3\times 10^4$
and also we considered the case in which the early dark energy era
started at $z=1100$ up to $z=10^4$. In both cases, the
amplification is significant, for $w=-0.9$ and for large reheating
temperatures. In conclusion, the scenario we described in this
paper indicates one certain thing: if a stochastic gravitational
wave signal is detected in future gravitational waves experiments,
the source of this signal is far from being certain, since several
scenarios might lead to such a spectrum.

\end{document}